%% file: sample-sigconf.tex
\documentclass[sigconf]{acmart}
\AtBeginDocument{%
  \providecommand\BibTeX{{%
    \normalfont B\kern-0.5em{\scshape i\kern-0.25em b}\kern-0.8em\TeX}}}

\usepackage{multirow}
\usepackage[linesnumbered,ruled,vlined]{algorithm2e}
\theoremstyle{definition}
\newtheorem{definition}{Definition}

\copyrightyear{2022} 
\acmYear{2022} 
\setcopyright{acmcopyright}\acmConference[DAC '22]{Proceedings of the 59th ACM/IEEE Design Automation Conference (DAC)}{July 10--14, 2022}{San Francisco, CA, USA}
\acmPrice{15.00}
\acmDOI{10.1145/3489517.3530522}
\acmISBN{978-1-4503-9142-9/22/07}

\makeatletter
\def\@copyrightspace{\relax}
\makeatother

\begin{document}

\title{Fast and Scalable Human Pose Estimation using mmWave Point Cloud}






 \author{Sizhe An}
 \affiliation{%
   \institution{University of Wisconsin-Madison}
   \city{Madison}
   \state{Wisconsin}
   \country{U.S.A.}}
  
 \author{Umit Y. Ogras}
 \affiliation{%
   \institution{University of Wisconsin-Madison}
   \city{Madison}
   \state{Wisconsin}
   \country{U.S.A.}}




\newcommand\todo[1]{{\color{red}#1}}
\newcommand\rev[1]{{\color{blue}#1}}
\newcommand*{\medcup}{\mathbin{\scalebox{1.25}{\ensuremath{\bigcup}}}}%

\begin{abstract}
Millimeter-Wave (mmWave) radar can enable high-resolution human pose estimation with low cost and computational requirements. 
However, mmWave data point cloud, the primary input to processing algorithms, is highly sparse and carries significantly less information than other alternatives such as video frames.
Furthermore, the scarce labeled mmWave data impedes the development of machine learning (ML) models that can generalize to unseen scenarios. We propose a fast and scalable human pose estimation (FUSE) framework that combines multi-frame representation and meta-learning to address these challenges. 
Experimental evaluations show that FUSE adapts to the unseen scenarios 4$\times$ faster than current supervised learning approaches and estimates human joint coordinates with about 7~cm mean absolute error.
\end{abstract}

\maketitle

\input{files/intro}

\input{files/related_work}

\input{files/proposed_framework}

\input{files/experiment}
\input{files/conclusion}

\begin{acks}
{This research was funded by NSF CAREER award CNS-2114499.}
\end{acks}
\vspace{-1mm}
\bibliographystyle{ACM-Reference-Format}
{\footnotesize{\bibliography{references/vision_ref,references/radar_ref, references/learning_ref}}}

\end{document}

%% file: files/intro.tex
\section{Introduction}

Human pose estimation refers to detecting and tracking key joints, such as wrists, elbows, and knees. 
It has rapidly growing applications areas, including rehabilitation, professional sports, and autonomous driving~\cite{simon-al-araji_2019, odemakinde_2021, vakanski2018data}. 
For example, one of the leading causes of autonomous car accidents is ``robotic'' driving, where the self-driver makes a legal but unexpected stop and causes other drivers to crash into it~\cite{odemakinde_2021}. 
Studies show that real-time human pose estimation can help computers understand and predict human motion, leading to more natural driving. 
Similarly, human pose estimation can enable remote rehabilitation applications, which are currently not feasible.

Human pose estimation can be performed by processing image, video, lidar (light detection and ranging), or mmWave radar data.  
The most common input type is RGB image and video frames since they offer accurate real-world representations using true color. However, the RGB frame quality depends heavily on the environmental setting, such as light condition and visibility. The lidar point cloud is a powerful alternative that overcomes these challenges. However, it has high cost and significant processing requirements, making them unsuitable for indoor applications such as rehabilitation. 
In contrast, mmWave radar can generate high-resolution 3D point clouds while maintaining low-cost and power advantages~\cite{ti_2020}. 

Using mmWave point cloud for human pose estimation faces two major challenges. 
First,  mmWave point cloud is significantly sparser and less informative compared to video and lidar point cloud data. 
For example, humans can easily recognize the object and its details from video and lidar point cloud, while it is almost impossible for people to interpret the mmWave point cloud representation accurately.
Second, the amount of labeled mmWave data lags severely behind video and lidar point cloud data. 
However, ML algorithms need a large amount of data to learn the generalization to new scenarios. 
For example, human pose estimation techniques must easily generalize to new users not included during training. 
\textit{Hence, there is a critical need for approaches that can perform well with fewer data points to harness the potential of mmWave data.}
This capability can enable home-based applications with significantly lower computation requirements since fewer data samples and training efforts are needed. 

Meta-learning has recently gained momentum because it can help ML models adapt to unseen scenarios faster with a few training iterations. 
It focuses on learning a strategy that generalizes to related yet unseen tasks from similar task distributions~\cite{finn2017model, biswas2018first}. 
It is first trained with a batch of tasks and learning rules designed to facilitate learning new tasks using only a few training iterations. 
In this way, the model employs the parameters sensitive to new samples, expediting generalization to new tasks.
The meta-learning concept fits the mmWave point cloud context very well since the amount of labeled training mmWave data is substantially smaller than video and lidar point cloud data. 
Hence, it can be a crucial enabler for mmWave radar point cloud-based human pose estimation with a few data samples and training iterations. 

This paper presents FUSE, a fast and scalable human pose estimation technique using mmWave point cloud. 
FUSE estimates the coordinates of 19 joints on the human body using mmWave point cloud as the input.
Its first component is a novel point cloud pre-processing method that fuses sparse frames to construct multi-frame data representations. 
Multi-frame representation enriches the sparse point cloud representation, reducing the mean absolute error~(MAE) on the human pose estimation task by 34\%, as demonstrated in Section 4.2. 
This method can easily be integrated into existing mmWave radar-based techniques as a pre-processing step to boost their performance without affecting other parts. 
The second component is a meta-learning framework that enables FUSE to adapt to the unseen data within a few epochs. 
Experimental results show that FUSE can converge to the optimal state in just five epochs, which is 4$\times$ faster than prior approaches. 

\vspace{1mm}
In summary, the major contributions of this paper are as follows:
\vspace{-5mm}
\begin{itemize}
    \item An effective point cloud pre-processing method that enhances mmWave point cloud representation by frame fusion,
    
    \item A meta-learning framework that significantly enhances the ability to generalize and adapt to unseen scenarios, 
    
    \item Experimental evaluations that show 7~cm MAE in estimating joint coordinates with 4$\times$ fewer training iterations than prior approaches (Our code is released for reproducibility.\footnote{https://github.com/SizheAn/FUSE})
    
    

    
    
\end{itemize}

In the rest, Section~\ref{sec: relatedwork} reviews the related work. Section~\ref{sec: proposed framework} introduces the mmWave background knowledge and the proposed approach. Section~\ref{sec:experiments} presents the experimental results. Finally, Section~\ref{sec: conclusions} concludes the paper.

%% file: files/related_work.tex
\section{Related Work}
\label{sec: relatedwork}

Early applications of mmWave radar focused on classification~\cite{singh2019radhar}, localization~\cite{lemic2016localization}, and obstacle detection~\cite{sugimoto2004obstacle} problems that do not require a high resolution. 
For example, a mmWave radar-based indoor human activity recognition technique is proposed in~\cite{singh2019radhar}. It recognizes five different activities: boxing, jumping, jumping jacks, squats, and walking with more than 90\% accuracy. 
Sugimoto et al.~\cite{sugimoto2004obstacle} present an obstacle detection method consisting of occupancy-grid representation and a segmentation method that divides the radar data. 
Similarly, Lemic et al.~\cite{lemic2016localization} propose a localization system that determines a mobile node's location using the flight time and arrival angles obtained by all the mmWave devices. 
In summary, these applications do not require a fine-grained representation due to their simplicity. 

Human pose estimation has attracted significant attention with recent advances in computer vision. It has a broad range of applications, including rehabilitation, professional sports, and autonomous driving~\cite{simon-al-araji_2019, odemakinde_2021, vakanski2018data}. 
Since it aims to reveal the nature of human motion (e.g., 3D joint coordinates), it requires a fine-grained representation. 
Zhao et al.~\cite{zhao2018rf} propose a technique that uses radio-frequency~(RF) antenna arrays reconstructs up to 14 body parts, including head, neck, shoulders, elbows, wrists, hip, knees, and feet. They compute 4D (time and three spatial axes) RF tensors 
using a 64-element antenna array with 60cm $\times$ 18cm area.
The massive customized antenna arrays enrich the input representation, but the large size and high cost significantly hinder practicality. 

Recent mmWave radar-based pose estimation techniques use point cloud representation from commercial radar devices like Texas Instrument~(TI) xWR1x43. Sengupta et al.~\cite{sengupta2020mm} propose mmPose, a human pose estimation technique that constructs the skeleton using mmWave point cloud and a forked-CNN architecture.  They use two radar devices and sum up the point values in the feature map level to overcome the sparse representation of the point cloud.
Xue et al.~\cite{xue2021mmmesh} present the mmMesh technique to construct human mesh using mmWave point cloud. 
It employs a human shape model
to strengthen the ability of deep learning models to predict human shape with fewer points. 
Finally, another recent study proposes a mmWave-based assistive rehabilitation system (MARS)~\cite{an2021mars} using human pose estimation. 
It sorts the mmWave point cloud and performs matrix transformations before feeding them to a CNN model. However, none of the prior techniques address two fundamental problems: mmWave point cloud data sparsity and the need for few-shot learning due to limited mmWave data.



The meta-learning concept was first proposed in~\cite{schmidhuber1987evolutionary}.
With the fast development of the deep learning industry, meta-learning has drawn significant attention again recently as its potential to reduce the training data requirements for the ML model. 
A few studies have recently applied meta-learning to point cloud~\cite{puri2020few, li2021few}. Puri et al. ~\cite{puri2020few} use MAML to solve the point cloud-based object classification and show it achieves similar accuracy with fewer data samples. Similarly, Li et al.~\cite{li2021few} propose few-shot meta-learning on point cloud for indoor semantic segmentation. However, both studies focus on the lidar point cloud, and their applications only involve simple classification and semantic segmentation.

In contrast to the previous studies, the proposed FUSE framework not only enhances the point cloud representation but also adapts to the unseen scenarios with a few training iterations. The novel point cloud pre-processing method using frame fusion increases the mmWave point cloud information by multiple times. In addition, \textit{for the first time}, we apply meta-learning in the field of mmWave point cloud for the challenging task-human pose estimation. 

%% file: files/proposed_framework.tex
\section{\hspace{-2.5mm}Proposed Pose Estimation Framework}
\label{sec: proposed framework}
\begin{figure}[b]
  \includegraphics[width=0.48\textwidth]{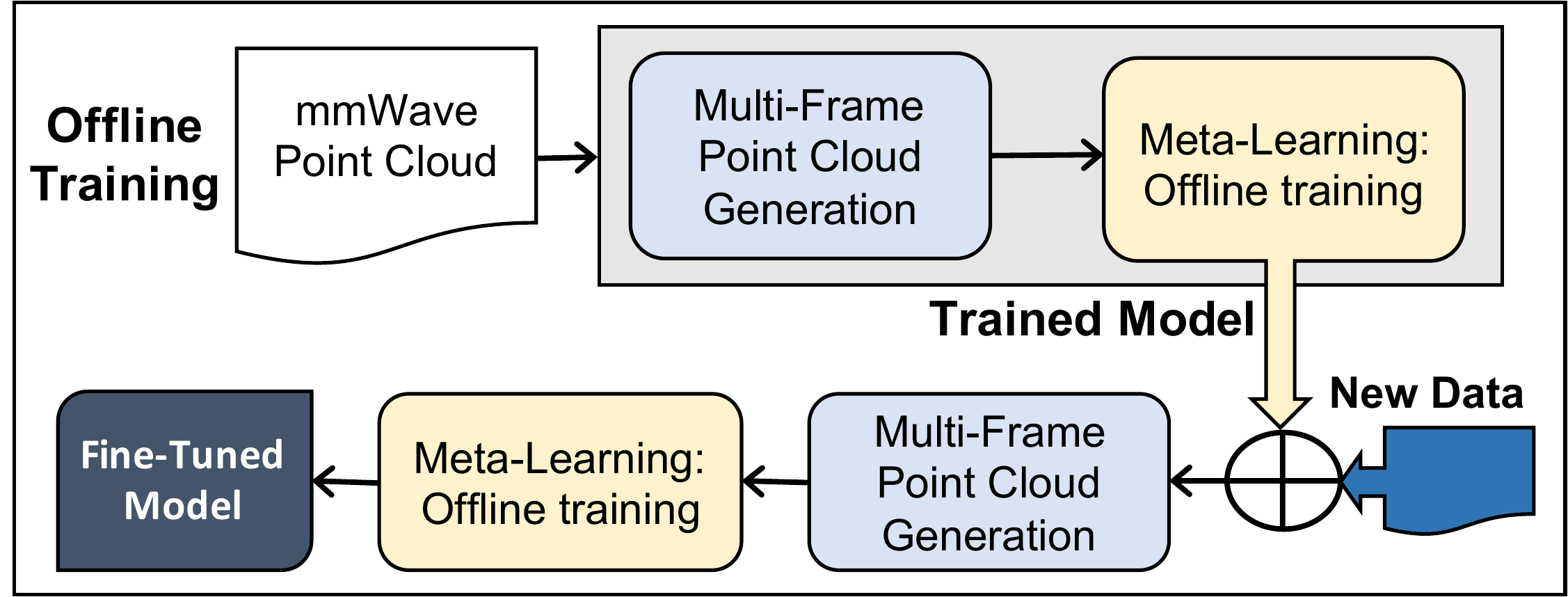}
  \caption{Overview of the proposed FUSE framework}
  \label{fig:overview}
\end{figure}

This section first provides a brief background on mmWave human pose estimation.
Then, it presents the proposed FUSE framework that consists of point cloud pre-processing and meta-learning steps, as described in Figure~\ref{fig:overview}.

\subsection{Background and Motivation}

\subsubsection{mmWave point cloud}
mmWave point cloud is usually generated by Frequency Modulated Continuous Wave (FMCW) radar using multiple transmit (Tx) and receiver antennas (Rx)~\cite{sengupta2020mm, zhao2018rf}. 
The radar transmits a chirp signal, a sinusoid wave whose frequency increases linearly with time, and receives the reflected signals at the RX antennas. 
Then, it processes the data to obtain range, velocity and angle resolutions using \textit{range FFT}, \textit{Doppler FFT}, and \textit{angle estimation} algorithms, respectively.
After eliminating noise with the constant false alarm rate~(CFAR) algorithm, the radar constructs a high-quality point cloud in the following format:
\begin{equation} \label{Eq:point}
P_i = \big(x_i, y_i, z_i, d_i, I_i \big), i \in \mathbb{Z}^+, 1 \leq i \leq N
\end{equation}
where $x_i$, $y_i$, $z_i$ are the spatial coordinates of the point, $d_i$ represents the Doppler velocity, $I_i$ denotes the signal intensity, and $N$ represents the total number of points in a given frame.

\subsubsection{Baseline convolution neural networks}
\label{sec: baselineCNN}
Convolutional neural networks~(CNNs) have become the mainstream method to process images and videos due to their ability to effectively extract feature maps from raw data~\cite{he2017mask}. 
Likewise, previous mmWave pose estimation studies~\cite{zhao2018rf, sengupta2020mm, an2021mars, xue2021mmmesh} 
employ CNNs to convert mmWave point clouds to 3D human joint coordinates. They typically use a few convolution layers with non-linear activation functions to extract the features from the input. Then, the intermediate feature maps from the convolution layers are flattened to one-dimensional feature vectors. Finally, fully-connected~(FC) layers process the one-dimensional feature vectors to produce the final output in the form of 3D coordinates of human joints. 
These networks are trained using ground truth joint locations obtained using a camera-based system. 
The mean absolute error (i.e., the $L1$ loss) between the ground truth and CNN predictions is used both during training and final evaluations.  

\subsubsection{Shortcomings of state-of-the-art techniques}
Prior mmWave pose estimation techniques focus on improving ML model accuracy for an existing set of users or movements. However, two critical aspects are ignored in these studies. First, they do not consider improving the sparse point cloud representation. 
For example, Zhao et al.~\cite{zhao2018rf} designed large antenna arrays to enrich the radar data. Similarly, Sengupta et al.~\cite{sengupta2020mm} use two mmWave radars and sum up their information. However, increasing the area or the number of antennas does not address the fundamental challenge. 
Instead, it increases the cost and makes deployment harder. Second, these studies rely on offline model training and testing. 
This choice is not practical since ML models need to quickly adapt to unseen scenarios in different application scenarios such as autonomous driving and rehabilitation. Thus, it is crucial to have an initial model that can fastly converge with any new data samples. To address these challenges, we propose a framework that enhances the point cloud representation and adapts to the unseen scenarios faster with a few training iterations.

\subsection{Multi-Frame Point Cloud Representation}
\label{sec: frames accumulation}
\begin{figure}[h]
  \includegraphics[width=0.5\textwidth]{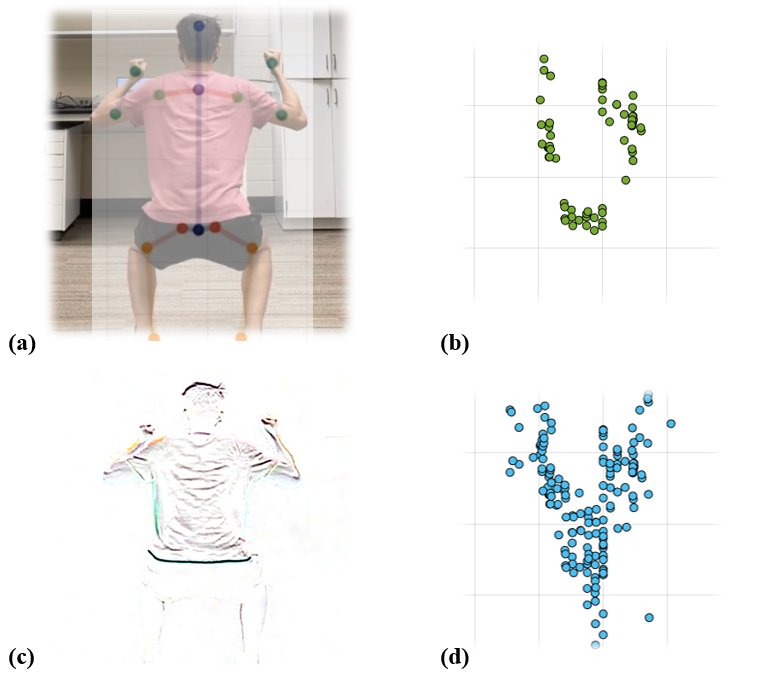}
  \caption{Visualization comparison of (a) one RGB frame, (b) a single-frame point cloud, (c) RGB residual frame, (d) proposed multi-frame point cloud.}
  \label{fig:multi_frame}
\end{figure}

As the sole data source to pose estimation, the point cloud must contain sufficient information to enable CNNs to extract the features, thus predicting accurate joint coordinates. 
However, current mmWave point cloud solutions only offer up to hundreds of points for one frame due to the limited number of antennas on the commercial mmWave radar~\cite{xue2021mmmesh}. 
For illustration, Figure~\ref{fig:multi_frame}(a) shows a video frame of a subject performing squat movement. With 512X424 frames (217K pixels), this frame can also be easily interpreted by humans. 
In strong contrast, the corresponding mmWave point cloud has only 64 3D points (192 data points) in a frame, i.e., almost 1000x fewer data points, as shown in Figure~\ref{fig:multi_frame}(b).
Consequently, it is harder for ML algorithms to extract information from this representation. 

The redundancy in video frames is often eliminated by using residual frames (i.e., differences between consecutive frames), which emphasize the changes due to motion~\cite{tao2020rethinking}. 
Indeed, Figure~\ref{fig:multi_frame}(c) illustrates that residual frames preserve the relevant information, facilitating feature extraction by ML algorithm. 
mmWave radar faces precisely the opposite problem: we must enrich a severely sparse representation as opposed to reducing redundancy. 
Therefore, \textit{for the first time in literature, we propose to fuse multiple sparse point cloud frames to synthesize a richer representation}. 
As illustrated by Figure~\ref{fig:multi_frame}(d), the proposed multi-frame approach significantly improve the interpretability compared to a single mmWave point cloud frame. 
Unlike a single-frame point cloud frame in Figure~\ref{fig:multi_frame}(b), 
multi-frame point cloud representation accurately captures the shape in the upper body. For example, we can see there are more points around the main body and arm area. 

Let $T_S > 0$ be the sampling period of the target mmWave radar (in this work, $T_S = 100~ms$).
The $k^{th}$ frame $f[k]$ in the point cloud contains the points collected during time interval $\big[kT_s, (k+1)T_s \big)$ for $k \in \mathbb{Z}^+$.  
Hence, we can express The $k^{th}$ frame $f[k]$ as:
\begin{equation}
\label{eq:frame}
    f[k] = \Big[ P_1[k], P_2[k], \ldots, P_N[k] \Big]~\forall k \in \mathbb{Z}^+    
\end{equation}
where $P_i[k]$ is the $i^{th}$ point in frame $k$ for $1 \leq i \leq N$.
Then, we fuse $M$ consecutive frames by concatenating them as follows:
\begin{equation}
\label{eq: fuse}
    F[k] = \Big\{ f[k-M], f[k-(M-1)], \ldots, f[k], \ldots, f[k+M-1], f[k+M] \Big\}
\end{equation}
where $M$ is a meta parameter that controls the number of fused frames. 
For instance, $M=1$ implies fusing three frames as $F[k] = \{f[k-1], f[k], f[k+1] \}$. 
This simple yet powerful idea significantly enhances the information content as interpretability even with $M=1$, as shown in Figure~\ref{fig:multi_frame}(d). 
Quantitative analysis in Section~\ref{sec: accumulation studies} 
demonstrates that our proposal can significantly improve the results of prior studies that employ mmWave point cloud data.  

\vspace{-3mm}
\subsection{Meta-Learning for Human Pose Estimation}

Meta-learning, also known as ``learning to learn",  aims at training a model on a variety of tasks such that it can solve new learning tasks using only ``few training samples". 
In our context, consider that a learning model, such as a CNN, is trained using an initial set of users and prescribed movements. 
If new users or movements are introduced, traditional techniques would either need to train models from scratch or adapt the model using incremental learning~\cite{an2020transfer,zhuang2020comprehensive}. 
Learning from scratch is highly inefficient, while the latter approach is an after-thought. 
\textit{In strong contrast, we construct the initial model by explicitly maximizing its ability to adapt to new users and movements using only a few training samples.} 
This capability is achieved by choosing model parameters sensitive to new samples, thereby ensuring that the gradient-based online learning rule can rapidly progress with new data. 
In this way, FUSE adapts to unseen scenarios quickly and estimates joint coordinates accurately after fine-tuning with a few training iterations.

\subsubsection{mmWave meta-learning setup}

This section defines the parameters and terminology used for meta-learning.

\begin{definition}[\textit{Training data}, $\mathcal{D}_{train}$]
\label{def:train_data}
The training data is the set of all fused frames $F[k], k \geq 1$ constructed using the point cloud frames as defined by Equation~\ref{eq: fuse}, i.e.,
\begin{equation} \label{eqn:training_data}
   \mathcal{D}_{train} = \medcup_{k}~F[k] 
\end{equation}
\end{definition}
%
\noindent Instead of directly using individual samples in  $\mathcal{D}_{train}$, meta-learning generates tasks and uses them for learning as described next. 
\begin{definition}[\textit{Task}, $\mathcal{T}$]
\label{def:train_data}
We define task $\mathcal{T}$ a set of fused frames uniformly sampled from the training data, i.e., $\mathcal{T} \sim \mathcal{D}_{train}$.
\end{definition}

Next, we present the proposed offline meta-training and online fine-tuning techniques.

\subsubsection{Offline Meta-Training}

After constructing the training data $\mathcal{D}_{train}$, 
we train the initial model using meta-learning using Algorithm~\ref{algo: Meta}.
The algorithm starts with randomly initializing the model parameters $\theta$. 
Then, it starts performing the meta-training iterations (lines 2--12). 
Each iteration $i$ starts with uniformly sampling a batch of tasks from the training data:
$\mathcal{T}_i \sim \mathcal{D}_{train}$ (line 3). 
Then, the inner loop (lines 4--10) operates on these tasks.
We first randomly choose a subset of tasks in the batch, and denote them as support tasks $\mathcal{T}^{sup}_i$.
The support tasks are used to update the intermediate model parameters
in each iteration using gradient descent:
%
\begin{equation} \label{Eq:update1}
\theta^{\prime}_i = \theta - \alpha\nabla_{\theta}L_{\mathcal{T}^{sup}_i}(g_\theta)
\end{equation}
where $\alpha$ is the $sample-level$ learning rate.
Our implementation uses the mean absolute error in joint coordinates (i.e., L1 distance) 
as the loss function, but other functions such as L2 can also be used (line 7). 
Unlike traditional supervised learning techniques, meta-learning samples next 
a set of query tasks $\mathcal{T}^{qry}_i$ (line 8) similar to the selection of the support tasks.
Then, the loss function for the query tasks are found 
for the model updated on line 7 (line 9).
\textit{After going over all tasks in $\mathcal{T}_i$}, 
the model updates its initial parameters $\theta$ using the summation of the loss on each query task $T^{qry}_i$, as shown in the following equation:
\begin{equation} \label{Eq:update2}
\theta = \theta - \beta\nabla_{\theta}\sum\limits_{\mathcal{T}^{qry}_i}L_{\mathcal{T}^{qry}_i}(g_{\theta^{\prime}_i})
\end{equation}

Note that the initial parameters $\theta$ are \textbf{only} updated once after all $T_i$ are done for a meta-training iteration. Also, in strong contrast to traditional transfer learning, 
these parameters
are updated using the \textit{intermediate parameters obtained from $\mathcal{T}^{sup}$} but the \textit{loss evaluated from $\mathcal{T}^{qry}$}.
In transfer learning, the initial parameters $\theta$ are updated using the intermediate parameters and the loss obtained from the same tasks. 
This crucial difference is why meta-learning can find the most sensitive parameters to the new data samples.


\begin{algorithm}[t]
\SetAlgoLined
\caption{Meta-training for mmWave point cloud}
\label{algo: Meta}
\KwIn{$\mathcal{D}_{train}$, $g_\theta$ (untrained model), $\beta$ (meta-learning rate)}
\KwOut{ML model that computes human joint coordinates using mmWave point cloud.}

Initialize the parameters $\theta$ of the ML model $g_\theta$

\For{each meta-training iteration}{
Sample a batch of tasks: $\mathcal{T}_i \sim \mathcal{D}_{train}$

\For{all $\mathcal{T}_i$ do}{
%
Sample support tasks from $\mathcal{T}_i$~: $\mathcal{T}^{sup}_i \subset \mathcal{T}_i$ \\
    Compute the gradient $\nabla_{\theta}L_{T^{sup}_i}(g_\theta)$ \\
    Update parameters $\theta^{\prime}_i = \theta - \alpha\nabla_{\theta}L_{T^{sup}_i}(g_\theta)$ \\
    Sample query tasks $\mathcal{T}^{qry}_i \subset \mathcal{T}_i$ \\
    Evaluate $L_{T^{qry}_i}(g_{\theta^{\prime}_i})$ using parameters $\theta^{\prime}_i$ \\
}

Update the initial parameters $\theta = \theta - \beta\nabla_{\theta}\sum\limits_{\mathcal{T}^{qry}_i}L_{\mathcal{T}^{qry}_i}(g_{\theta^{\prime}_i})$
}

\end{algorithm}


\subsubsection{Online Fine-Tuning Phase}
\label{sec: finetune}

So far, we presented the construction of the initial meta-learned model using the available training data. 
Suppose a new user or movement, which is absent from the training data, 
is introduced in the field, i.e., after the trained model is deployed. 
Our goal is to use few training samples denoted by $\mathcal{D}_{test}$ 
to update the initial model. 
We emphasize that the cardinality of the test data is 
$|\mathcal{D}_{test}| << |\mathcal{D}_{train}|$. 
Section~\ref{sec: faster adaptation} validates this claim and 
shows that the size of the required test data is 
4$\times$ smaller than that required by supervised techniques.

We use $\mathcal{D}_{test}$ to perform the fine-tuning and testing to evaluate our meta-trained model's performance. The model takes a part of $\mathcal{D}_{test}$ to perform forward pass and back-propagation to fine-tune. Then, we use the other part of $\mathcal{D}_{test}$ only to perform inference and evaluate the model. The exact procedure to split the data is described with the implementation in Section~\ref{sec: dataset}. Ideally, only fine-tuning for a few iterations should be enough since the model learns the generalization of the point cloud. In summary, the fine-tuning phase does not require any extra steps, facilitating online usage.

%% file: files/experiment.tex
\section{Experimental Evaluations}
\label{sec:experiments}

\subsection{Experimental Setup and Baseline Model}
\label{sec: dataset}

\noindent\textbf{Human Pose Estimation Data:}
We employ an open-source mm-Wave point cloud dataset (MARS~\cite{an2021mars}) to evaluate the proposed FUSE framework. The dataset consists of 40,083 labeled frames (defined in Equation~\ref{eq:frame}) collected using TI IWR1443 Boost mmWave radar~\cite{ti_2020}. 
The frames correspond to 10 distinct rehabilitation movements performed by four human subjects in front of the mmWave radar and Microsoft Kinect V2 sensor. 
The reference coordinates of 19 joints are found using Kinect V2 and added as labels to the mmWave data at a 10~Hz sampling rate. 
Then, each movement data is individually split into 60\% training, 20\% validation, and 20\% test sets. 

\noindent\textbf{Baseline ML model:}
We implement the CNN trained with the MARS dataset~\cite{an2021mars} as the baseline model to ensure a fair comparison. 
It has two convolution layers with Rectified Linear Unit~(ReLU) activations, followed by two FC layers, with a total model of 1,095,115 parameters.
The number of neurons of two FC layers is 512 and 57, respectively. Here, the output values of the final 57 neurons represent 19 human joints coordinates values in $x$, $y$, $z$-axes. 
The proposed CNN trained using the FUSE framework has the same dimensions and model size for a fair comparison.
%

\noindent\textbf{Implementation details:}
We implemented all baseline and meta-learning approaches using PyTorch 1.8.1 ~\cite{paszke_2019}. The training and testing are performed on a Nvidia GeForce RTX 3090 graphic card with 24GB of memory.
The meta-learning approach is based on MAML-PyTorch implementation~\cite{MAML_Pytorch}. 
Our results can be reproduced using the following hyper-parameter values: 
20,000 meta-training iterations, 
32 tasks sampled for each iteration,  
\textit{sample-level} learning rate $\alpha =$ 0.1, 
\textit{task-level} meta-learning rate $\beta =$ 0.001. 
During meta-training, each support task and query task $\mathcal{T}_i$ samples 1,000 frames from $\mathcal{D}_{train}$ randomly. 
We use 200 frames from $\mathcal{D}_{test}$ to fine-tune and the all rest frames from $\mathcal{D}_{test}$ to evaluate the performance. 
Finally, MAE~(i.e. the L1 loss) loss function and Adam optimizer~\cite{DBLP:journals/corr/KingmaB14} are employed for calculating the loss and updating the gradients. 



\subsection{Multi-Frame Fusion of Point Cloud Data}
\label{sec: accumulation studies}
Fusing multiple frames enriches the information content of the mmWave point. 
To study the impact of frame the fusion alone, this section uses the baseline CNN architecture, 
the default 60\% - 20\% - 20\% data split, 
and training parameters (128 batch size and 150 epochs).
We conduct experiments on three settings: single-frame (baseline), fuse three frames, and fuse five frames. 

Table~\ref{tab: frames accumulation} summarizes the average MAE in predicting the joint coordinates with and without multi-frame fusion. 
Without changing any other parameters, 
fusing three frames \textit{consistently decreases the average MAE} along $x-$, $y-$, and $z-$axis, 
and average MAE reduction from 5.5~cm to 3.6~cm. 
Fusing more frames does not continuously improve the accuracy since redundancy is introduced. 
Specifically, we observe that fusing three frames outperforms a single frame by 1.9~cm margin (34\%). 
This improvement is impressive since achieving similar savings without frame fusion would require a significant 
increase in the model complexity. 

These controlled experiments show that fusing multiple frames enhances the point cloud representation, thus improving the performance of human pose estimation. 
Hence, it can boost the performance of existing mmWave radar techniques without affecting the ML models they employ. In the rest, we fuse three frames since it leads to significant savings with negligible overhead.


\vspace{-2mm}
\begin{table}[h]
\caption{MAE of the baseline model under different frame fusion settings.} 
\label{tab: frames accumulation}
\vspace{-4mm}
\begin{tabular}{ccccc}
\toprule
                                                            & X (cm) & Y (cm) & Z (cm) & Average (cm) \\ \cmidrule{2-5}
Single-frame                                                & 6.4                   & 3.6              & 6.5                 & 5.5            \\ 
\begin{tabular}[c]{@{}c@{}} Fuse 3 Frames\end{tabular} & 4.2                   & 2.5              & 4.4                 & 3.6            \\ 
\begin{tabular}[c]{@{}c@{}} Fuse 5 Frames\end{tabular} & 6.9                   & 4.1              & 5.5                 & 5.5            \\ \bottomrule
\end{tabular}
\end{table}
\vspace{-5mm}

\subsection{\hspace{-1mm}Convergence Time and Accuracy Evaluation}
\label{sec: faster adaptation}

\subsubsection{Data splitting}
To examine the ability of FUSE to adapt to new scenarios, 
this experiment splits the dataset to capture the worst-case scenario.
The training and validation sets \textit{exclude all data} from one particular movement (``\textit{right limb extension}") and one of the users (user 4).  
With this split, the test data ($\mathcal{D}_{test}$) seen only during fine-tuning has only 749 frames, justifying our claim about few samples available online. 
In contrast, the training data ($\mathcal{D}_{train}$) consists of 29,225 frames from the remaining nine movements and users. 
A more comprehensive leave-one-out experiment is left for future work due to space and execution time considerations.


\subsubsection{Quantitative results}

Fine-tuning is a commonly used method in transfer learning~\cite{zhuang2020comprehensive} to test a model's ability to adapt to new data samples.
It fine-tunes all layers or part of a pre-trained model with new data samples~\cite{zhuang2020comprehensive}. 
We conduct experiments for both cases: fine-tuning all layers and only the last FC layer with its activation. 

\begin{figure}[h]
  \includegraphics[width=0.48\textwidth]{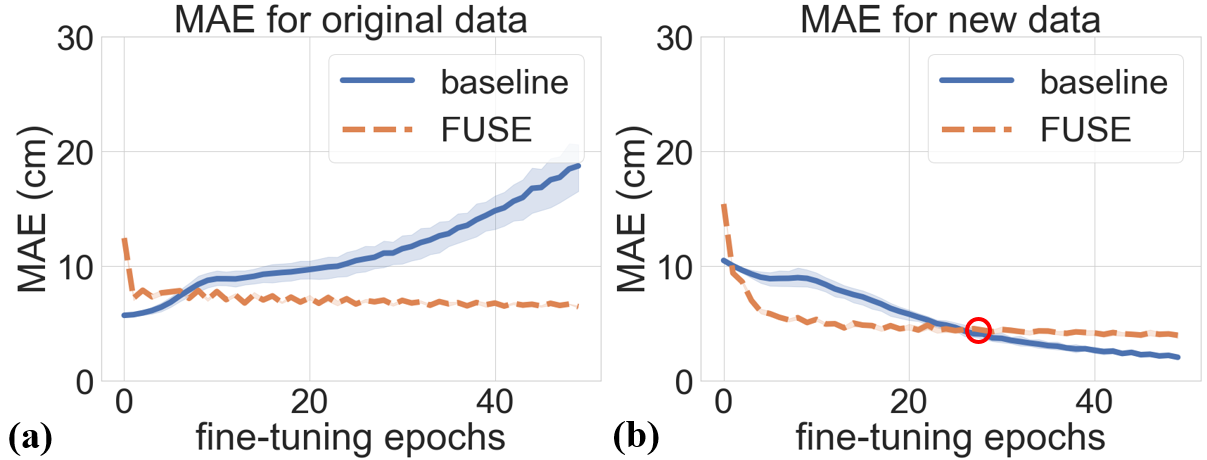}
  \caption{MAE comparison between baseline and FUSE model for fine-tuning all layers.}

  \label{fig:MAE_alllayers}
  \vspace{-3mm}
\end{figure}

\noindent\textbf{Fine-tune \textit{all layers}: }
Figure~\ref{fig:MAE_alllayers} shows the MAE comparison between baseline and FUSE model fine-tuned for all layers.
The baseline model achieves a remarkable MAE of 6.7~cm after the initial training with the original data available offline, as shown in Figure~\ref{fig:MAE_alllayers}(a).
In contrast, the proposed FUSE model starts with 12.4~cm MAE since it is optimized for generalization rather than fitting to known cases. 
Indeed, FUSE achieves almost 6.0~cm MAE 
after only 5 fine-tuning epochs with the new data. 
\textit{We emphasize that both the baseline model and FUSE are fine-tuned using 
the new data, which is not included in the initial training.}
The extra data improves FUSE's performance even on the original training dataset. 
After fine-tuning, the MAE of FUSE stabilizes at about 9.4~cm for the original data (Figure~\ref{fig:MAE_alllayers}(a)) and 
4.0~cm for the new user data (Figure~\ref{fig:MAE_alllayers}(b)).

Figure~\ref{fig:MAE_alllayers}(b) also shows that the baseline model can be effectively fine-tuned for the new data. 
In strong contrast to FUSE, the improved performance comes at a steep penalty for the original data, as shown by the solid line in Figure~\ref{fig:MAE_alllayers}(b). 
As the baseline model adapts to new data, it forgets the original one, implying that it tends to overfit rather than learn the trends.


In summary, FUSE is able to achieve 6.0~cm MAE, $\sim$3~cm lower than the baseline with only 5 epochs fine-tuning.
With 26 epochs, as a red circle shown in Figure~\ref{fig:MAE_alllayers}(b), baseline approach is able to achieve 4.6~cm, which is comparable to 4.3~cm for FUSE model. However, it is at the expense of forgetting the original data. 
The MAE for original data reaches 10.6~cm, as summarized in Table~\ref{tab: MAEcompare} (columns labeled ``All layers"), since the baseline model do not learn the generalization. After that, the baseline approach just keep memorizing the new data and forgetting the original data.

\begin{table}[h]
\setlength{\tabcolsep}{5pt}
\renewcommand{\arraystretch}{0.8}
\caption{MAE comparison between baseline and FUSE model. Results of both fine-tuning all layers and only the last layer are presented in the table. Intersection means the epoch where baseline's MAE meets FUSE's for the new data. It is 26 epochs for all layers and 16 epochs for the last layer 
(marked by a red circle in the second plot.}
\vspace{-3mm}
\label{tab: MAEcompare}
\begin{tabular}{lcrrrr}
\toprule
 &
  \multicolumn{1}{l}{} &
  \multicolumn{2}{c}{All layers} &
  \multicolumn{2}{c}{Last layer} \\ \cmidrule{3-6} 
\multicolumn{1}{c}{}       &          & baseline & FUSE  & baseline & FUSE  \\ \midrule 
\multirow{2}{*}{5 epochs}  & Original & 6.4    & 7.6  & 6.5   & 9.0  \\ \cmidrule{2-2}
                           & New      & 9.0    & 6.0  & 9.6    & 8.3  \\ \midrule 
\multirow{2}{*}{\begin{tabular}[c]{@{}l@{}} Intersection \end{tabular}} 
                            &  Original &  10.6 &  6.6 &  7.2 &  8.2 \\ \cmidrule{2-2}
                           & New      & 4.6     & 4.3  & 7.1     & 7.0  \\ \midrule 
\multirow{2}{*}{50 epochs} & Original & 18.7    & 6.4 & 31.0    & 7.8 \\ \cmidrule{2-2}
                           & New      & 2.0     & 3.9  & 3.9     & 6.0  \\ \bottomrule
\end{tabular}
\end{table}

\begin{figure}[t]
  \includegraphics[width=0.49\textwidth]{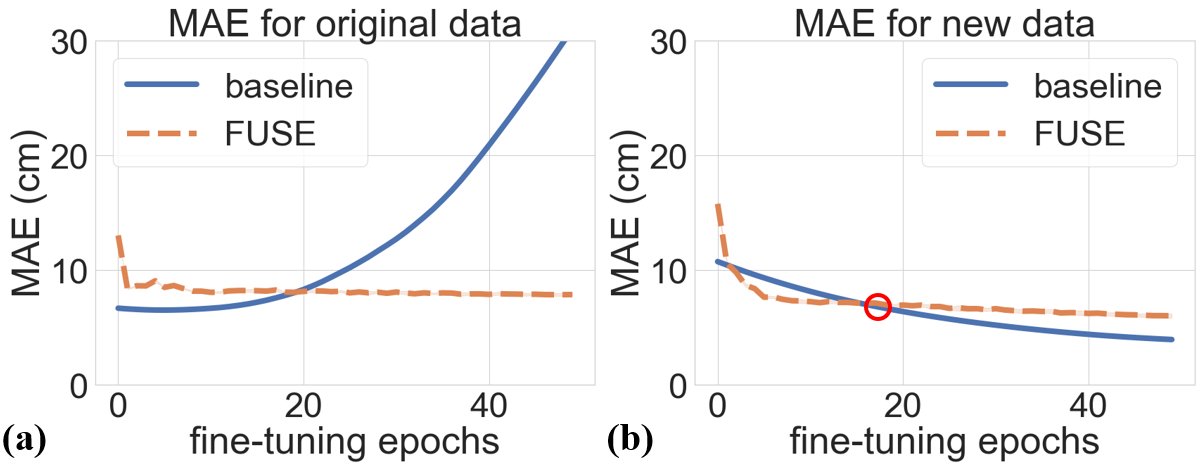}
  \caption{MAE comparison between baseline and FUSE model for fine-tuning the last layer.}
  \vspace{-5mm}
  \label{fig:MAE_lastlayers}
\end{figure}

\noindent\textbf{Fine-tune \textit{the last layer}: }
Figure~\ref{fig:MAE_lastlayers} shows the MAE comparison between the baseline and FUSE model when only the last fully connected layer is fine-tuned. 
It shows a very similar pattern with fine-tuning with all layers, as summarized in Table~\ref{tab: MAEcompare} (columns labeled ``last layer"). 
With only 5 epochs of fine-tuning, the FUSE model achieves 8.3~cm MAE, 1.3~cm lower than the baseline. With 16 epochs, as shown with a red circle in Figure~\ref{fig:MAE_lastlayers}(b), the baseline approach can achieve 7.1~cm, which is comparable to 7.0~cm for FUSE model. However, it is at the expense of forgetting the original data, as the MAE for original data reaches 7.2~cm. The baseline approach memorizes the new data and forgets the original data to adapt to new scenarios. Compared to fine-tuning all layers, fine-tuning only the last layer yields a higher error for new data and significantly worse forgetting trend. The results show that fine-tuning with all layers can help the model adapt to unseen scenarios since FUSE learns the generalization.

In summary, FUSE enhances the mmWave point cloud representation, thus improving the human pose estimation performance significantly. In addition, FUSE adapts to the unseen data within five epochs. This is 4$\times$ faster than the baseline approach that needs at least 20 epochs to achieve the same performance, at the expense of forgetting the original data.

\vspace{-2mm}

%% file: files/conclusion.tex
\section{Conclusions}
\label{sec: conclusions}

Human pose estimation offers valuable insight for many applications, including rehabilitation, professional sports, and autonomous driving. mmWave radar-based pose estimation is emerging as a promising direction due to low cost, power consumption, computation requirements. However, it suffers from sparsity representation and scarce labeled data. 
This paper presented  FUSE, a mmWave point cloud-based human pose estimation framework, to mitigate these challenges. FUSE consists of a novel multi-frame pre-processing method and meta-learning components. Multi-frame representation alone reduces the MAE of joint coordinate estimates by 34\% compared to prior work. In addition, the meta-learning enables the ML model to adapt to the unseen scenarios within five epochs, 4$\times$ faster than existing approaches.
\vspace{-1mm}